# Multifractal scaling analyses of the spatial diffusion pattern of COVID-19 pandemic in Chinese mainland

Yuqing Long, Yanguang Chen, Yajing Li

(Department of Geography, College of Urban and Environmental Sciences, Peking University, Beijing 100871, P.R.China. E-mail: chenyg@pku.edu.cn)

**Abstract:** Revealing spatiotemporal evolution regularity in the spatial diffusion of epidemics is helpful for preventing and controlling the spread of epidemics. Based on the real-time COVID-19 datasets by prefecture-level cities, this paper is devoted to exploring the multifractal scaling in spatial diffusion pattern of COVID-19 pandemic and its evolution characteristics in Chinese mainland. The ArcGIS technology and box-counting method are employed to extract spatial data and the least square regression based on rescaling probability ($\mu$-weight method) is used to calculate fractal parameters. The results show multifractal distribution of COVID-19 pandemic in China. The generalized correlation dimension spectrums are inverse S-shaped curves, but the fractal dimension values significantly exceed the Euclidean dimension of embedding space when moment order $q<<0$. The local singularity spectrums are asymmetric unimodal curves, which slant to right. The fractal dimension growth curves are shown as quasi S-shaped curves. From these spectrums and growth curves, the main conclusions can be drawn as follows: First, self-similar patterns developed in the process of COVID-19 pandemic, which seem be dominated by multifractal scaling law. Second, the spatial pattern of COVID-19 across China can be characterized by global clustering with local disordered diffusion. Third, the spatial diffusion process of COVID-19 in China experienced four stages, i.e., initial stage, the rapid diffusion stage, the hierarchical diffusion stage, and finally the contraction stage. This study suggests that multifractal theory can be utilized to characterize spatio-temporal diffusion of COVID-19 pandemic, and the case analyses may be instructive for further exploring natural laws of spatial diffusion.

**Keywords:** COVID-19; spatial diffusion; spatial isolation; multifractals; fractal dimension growth curve; Chinese mainland

# 1. Introduction

Human beings have long been suffered from various rapidly spread epidemics. It is significant to research the spatial diffusion regularity of infectious diseases. The most recent case of epidemic outbreak is the well-known coronavirus (COVID-19) in December 2019 (Huang *et al*., 2020; Mehta *et al*., 2020). Studies on the spatial diffusion pattern of COVID-19 are not only helpful for understanding its transmission dynamics, but also for the future prevention and control of epidemics. In literature, many studies have analyzed and modeled the spatial diffusion pattern of epidemics by various mathematical methods (Chen *et al*, 2021; Fang *et al*., 2009; Kang *et al*., 2020; Meng *et al*., 2005; Wang *et al*., 2008; Wu *et al*., 2020; YeŞİLkanat, 2020; Zhao and Chen, 2020). Conventional mathematical methods are based on the mathematical concept of characteristic scale, which is always termed characteristic length in literature (Hao, 1986; Liu and Liu, 1993; Takayasu, 1990; Wang and Li, 1996). If a spatial distribution bears no characteristic length, conventional mathematical modeling will be ineffective. In this case, the modeling idea from characteristic scale should be replaced by that from scaling. Fractal geometry provides a powerful tool for scaling analysis (Mandelbrot, 1982), and has been widely used to characterize complex systems such as cities (Batty and Longley, 1994; Chen, 2014a; Eke *et al*., 2002; Frankhauser, 1998). In this study, we propose a multifractal scaling model to capture the complex pattern and process of spatial epidemic dynamics.

Multifractal scaling can be employed to quantitatively describe various spatial heterogeneus phenomena. In recent years, growing studies have employed multifractal scaling modeling to geospatial pattern analysis (Chen and Wang, 2013; Frankhauser *et al*., 2018; Man *et al*., 2019; Murcio *et al*., 2015; Salat *et al*., 2018). Multifractal measures reflect a distribution of physical or other quantities on a geometric support (Feder, 1988). The spatial diffusion pattern of the epidemic is also the reflection of spatial organization patterns and cascade system of human social and economic activities (Wang *et al*., 2020). Previous studies have found that the spatial distribution of population exhibits multifractal structure (Appleby, 1996; Semecurbe *et al*., 2016; Liu and Liu, 1993), and the epidemic infections spread among the population. Population may be the geometric support of multifractal distribution of COVID -19 pandemic. It is valuable to figure out whether the spatial diffusion pattern of COVID-19 shows multifractal characteristics, and what can be learned from the spatio-temporal analysis of multifractal scaling model for spatial diffusion.

To address the above problems, we perform the multifractal scaling analyses to quantify the spatial diffusion patterns of COVID-19 pandemic. The multifractal measures are applied to spatial network datasets rather time series, but the fractal parameters form a set of sample paths, which can be used for simple time series analysis. The evolution characteristics of COVID-19 pandemic diffusion in Chinese mainland can be revealed by the sample paths. The analytical process is as below: First of all, a spatial database of COVID-19 infections by prefecture-level cities on different dates is established. Then, the functional box-counting method and the ordinary least square (OLS) regression based on rescaled probability are employed to calculate multifractal parameters. Finally, multifractal-based spatial analysis is made. The remainder of this paper is organized as follows. In Section 2, the multifractal scaling models are introduced, and dataset and analytical procedure used in this study is illuminated. In Section 3, the main results of multifractal analyses are presented. The multifractal characteristics of the spatial pattern of COVID-19 are examined, and then two sets of multifractal spectrums are employed to reveal detailed characteristics of the spatial diffusion pattern. Finally, the spatial diffusion process and its stage features are illustrated by the fractal dimension growth curves. In Section 4, the key points of the analyzed results are outlined, and related questions are discussed. Finally, in Section 5, the discussion is ended by drawing the main conclusions of the spatial diffusion pattern of COVID-19 pandemic in Chinese mainland.

## 2. Methodology and materials

### 2.1 Multifractal model of multifractal system

Multifractal scaling analysis bears analogy with telescopes and microscopes in spatial analysis. It provides two sets of parameter spectrums adequately quantifying the spatial patterns and the statistical distribution of measurements across spatial scales (Pavon-Dominguez *et al*., 2017; Stanley and Meakin, 1988). The spatial pattern of COVID-19 pandemic in Chinese mainland can be treated as a spatial heterogeneous system, and multifractal measures can be employed to characterize its complexity. A multifractal system is a self-similar hierarchy with cascade structure, which is based on two or more scaling processes (Chen, 2014a) (Figure 1(a)). Multifractals are also known as multi-scaling fractals. Therefore, different subareas have different scaling behaviors and growth probabilities, i.e., the denser and sparser zones may display different spatial characteristics, leading to a heterogeneous distribution. Therefore, a single fractal dimension is difficult to fully describe the patterns and processes of complex systems in the real world (Chen, 2014a; Murcio *et al*., 2015). The multifractal analyses provide a series of parameters to capture detailed information about the rich structure at fine scales, which finally appear in a multifractal spectrum of parameters (Caniego *et al*., 2005) (Figure 1(b)). Generally, two sets of parameters are employed to make multifractal analyses, including global and local parameters. The global parameters consist of *generalized correlation dimension*, $D(q)$, and *mass exponent*, $\tau(q)$, which reveal overall characteristics of multifractal systems. The local parameters comprise *singularity exponent*, $\alpha(q)$, and *local fractal dimension* of the fractal subsets, $f(\alpha)$, which reveal local characteristics of multifractal structure (Feder, 1988). The parameter $q$ refers to the moment order ($-\infty<q<\infty$), and by changing the value of $q$, the fractal characteristics of the spatial pattern of COVID-19 with different densities can be reflected. In the multifractal spectra, when $q\to\infty$, attention can be focused on locations with high density. Corresponding to the geographical space, it represents the spatial characteristics of the core area with more COVID-19 infections; when $q\to-\infty$, attention can be focused on locations with low density. Corresponding to the geographical space, it represents the spatial characteristics of the sparse regions with few COVID-19 infections (Figure 1).

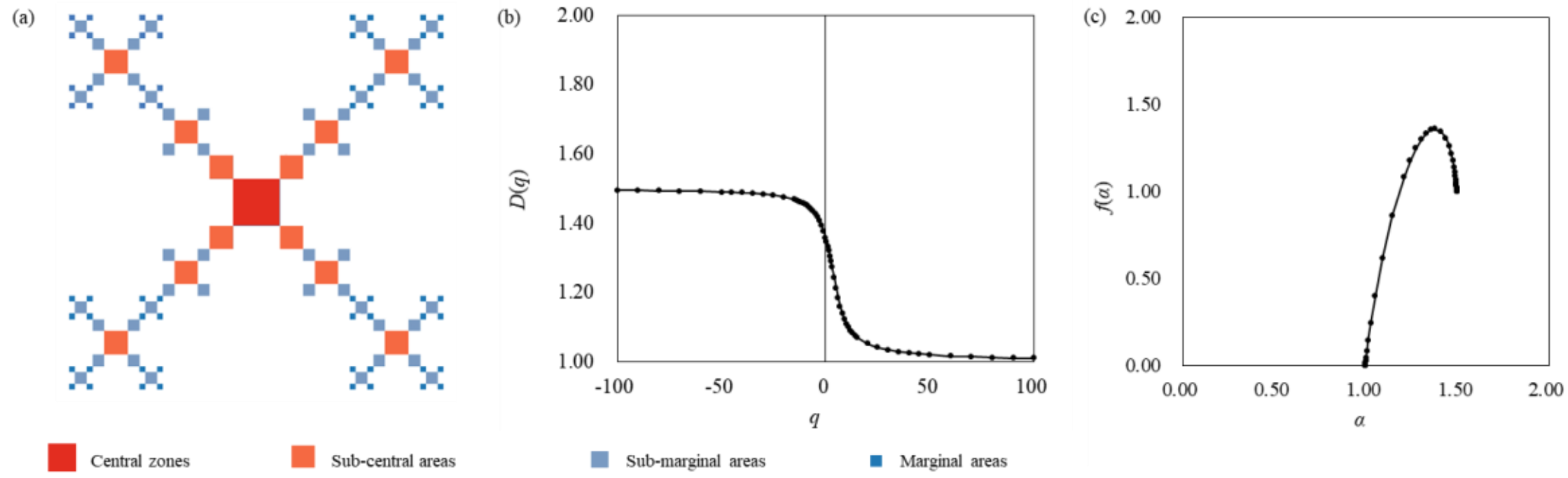

**Figure 1. Schematic illustration of how different dominant structures of systems are projected to the multifractal spectrums.** (a) Different hierarchical structures of regular multifractals. (b) The *generalized correlation dimension* spectrum, $D(q)$ curve, and singularity spectrum $f(\alpha)$ curve of regular multifractals.

The diffusion pattern of COVID-19 in Chinese Mainland can be seen as a spatial heterogeneous network, consisting of various cities as nodes. For a given date, the confirmed cases of COVID-19 form various sizes of clusters around urban nodes. Geographical location, urban population size, and government control means can affect the chances of case occurrence. Different opportunities for virus transmission manifest as different probabilities of case occurrence, and different probabilities reflect different case distribution densities. So, two measures can be used to describe the network structure of virus diffusion at given time. One is the growth probability of COVID-19 cases, $P_i$, and the other is the linear size of case clusters, $\varepsilon_i$, where $i$=1, 2, 3, …, $N$ represents the number of case clusters. If the transmission and distribution network of COVID-19 has fractal properties, there is a scaling relationship between the linear size $\varepsilon_i$ and growth probability $P_i$, and this suggests that the probability depends on linear size. So, the clusters represents different fractal units, which form a self-similar hierarchy with cascade structure. If the fractal structure bear multiple scaling processes, then we have

$$\sum_{i=1}^{N} P_i(\varepsilon)^q \varepsilon_i^{(1-q)D(q)} = \sum_{i=1}^{N} P_i(\varepsilon)^q r_i^{-\tau(q)} = 1, \tag{1}$$

which is a transcendental equation characterizing varied systems. In many cases, it is neither necessary nor possible for us to describe the distribution clusters of cases of various sizes separately. A smart approach is to transform the probability distribution reflecting density differences into multifractal parameter spectrums by means of unified linear size, $\varepsilon$, where $\varepsilon \equiv \varepsilon_i$. A desirable approach is box-counting method, and thus $\varepsilon$ indicates the linear size of box. Based on box-counting method, equation (1) changes to

$$\sum_{i=1}^{N} P_i(\varepsilon)^q = \varepsilon^{\tau(q)} = \varepsilon^{(q-1)D(q)}, \tag{2}$$

which is the basic expression of Renyi entropy and global parameter of multifractals.

Global parameters describe the research object such as network of COVID-19 diffusion from the macro level. The generalized correlation dimension $D(q)$ is defined on the basis of Renyi's entropy (Hentschel and Procaccia, 1983; Feder, 1988; Vicsek, 1989). Considering l'Hospital's rule, the formula of generalized correlation dimension can be derived from equation (2) as follows

$$D(q)=-\lim_{\varepsilon\to 0}\frac{I_q(\varepsilon)}{\ln\varepsilon}=\begin{cases}\dfrac{1}{q-1}\lim\limits_{\varepsilon\to 0}\dfrac{\ln\sum_{i=1}^{N(\varepsilon)}P_i(\varepsilon)^q}{\ln\varepsilon}, & (q\neq 1)\\ \lim\limits_{\varepsilon\to 0}\dfrac{\sum_{i=1}^{N(\varepsilon)}P_i(\varepsilon)\ln P_i(\varepsilon)}{\ln\varepsilon}, & (q=1)\end{cases}, \tag{3}$$

where $q$ refers to the moment order ($-\infty<q<\infty$), $I_q(\varepsilon)$ to the Renyi's entropy with a linear size $\varepsilon$, and $P_i(\varepsilon)$ is the growth probability of the $i$th fractal unit. Based on box-counting method, a set of fractal units at given level are replaced by nonempty boxes with corresponding linear scales $\varepsilon$. Therefore, when measured by box-counting method, $N(\varepsilon)$ refers to the number of nonempty boxes, and $P_i(\varepsilon)$ represents the growth probability in the $i$th box, indicating the distribution probability of COVID-19 infections . At a specific scale $\varepsilon$, the larger the $P_i(\varepsilon)$, the higher the density of COVID-19 infections, which corresponds to central regions. According to equation (2), another global parameter, mass exponent $\tau(q)$, can be estimated by $D(q)$, and the formula is (Halsey *et al*., 1986; Feder, 1988)

$$\tau(q)=(q-1)D(q), \tag{4}$$

which reflects the properties from the viewpoint of mass. The global analysis relies heavily on the generalized correlation dimension $D(q)$. The changing curve of $D(q)$ with $q$ forms the global multifractal spectrum.

Local parameters focus on the micro-level in a multifractal network of COVID-19 diffusion. Due to its heterogeneity, the multifractal system has many fractal subsets, taking on various cluster of case distribution. As indicated above, there is a scaling relation between the growth probability of COVID-19 and corresponding linear size of cluster of cases. This relation suggests a power law as below

$$P_i(\varepsilon)\propto\varepsilon_i^{\alpha(\varepsilon)}, \tag{5}$$

where $\varepsilon_i$ refers to the corresponding linear size of the $i$th fractal unit, and $\alpha(q)$ denotes the strength of local singularity, also known as *Lipschitz-Hölder singularity exponent*, suggesting the degree of singular interval measures (Feder, 1988). Different values of $\alpha$ correspond to different subsets of multifractals. Accordingly, the number of fractal subsets with the same $\alpha$ value under the linear $\varepsilon_i$ is given by

$$N(\alpha,\varepsilon_i) \propto \varepsilon_i^{-f(\alpha)}, \tag{6}$$

where $f(\alpha)$ refers to the *local fractal dimension*. Therefore, the different subset will also have a corresponding fractal dimension, composing a singularity spectrum $f(\alpha)$ that changes along with $\alpha(q)$ to describe the corresponding changes in the subsystem (Song and Yu, 2019). The curve of correlation between $f(\alpha)$ and $\alpha(q)$ forms the local multifractal spectrum. The global parameters and local parameters of multifractals can be associated with Legendre transform, that is

$$\alpha(q) = \frac{\mathrm{d}\tau(q)}{\mathrm{d}q} = D(q) + (q-1)\frac{\mathrm{d}D(q)}{\mathrm{d}q}, \tag{7}$$

$$f(\alpha) = q\alpha(q) - \tau(q). \tag{8}$$

If we use a certain algorithm to calculate the global parameters, $D(q)$ and $\tau(q)$, we can obtain the local parameters, $\alpha(q)$ or $f(\alpha)$, by means of equations (7) and (8), and vice versa.

As mentioned earlier, the definition of multifractal dimension spectrum is based on Renyi entropy and generalized correlation function. With the help of entropy, the spatial difference and space-filling pattern of COVID-19 can be characterized. With the help of correlation function, the diffusion mechanism and controlled degree of COVID-19 can be revealed. In this study, the main measure indexes are as follows: the generalized correlation dimension spectrum $D(q)$, the local singularity spectrum $f(\alpha)$ and three representative fractal dimensions: the *capacity dimension* $D_0$, the *information dimension* (*Shannon entropy*) $D_1$, and the *correlation dimension* $D_2$ (Grassberger, 1983). First of all, the log-log plots of $D_0$, $D_1$ and $D_2$ are drawn to reveal the fractal feature of the spatial pattern of COVID-19. Secondly, the value of $D_0$, $D_1$ and $D_2$ are compared to determine whether it can be characterized by multifractal structure (Chen, 2014; Murcio *et al*., 2015). Thirdly, the global multifractal spectrum $D(q)$ is employed to reflect overall characteristics. In theory, based on box-counting methods, the multifractal dimension values should come between topological dimension $d_T$=0 and embedding dimension $d_E$=2 (Huang and Chen, 2018). Fourthly, the local singularity spectrum, i.e., the $f(\alpha)$ curve, is employed to reflect local characteristics. In theory, if it appears as a right-leaning unimodal curve, the fractal growth is dominated by spatial concentration. Conversely, if it appears as a left-leaning unimodal curve, this suggests the fractal pattern of spatial deconcentration (Chen, 2014a). Lastly, the time varying curves of $D_0$, $D_1$ and $D_2$ are drawn to demonstrate the stage characteristics of spatial diffusion of COVID-19.

## 2.2 Data sources and methods

The main aim of this study is at the complex structural characteristics of the spatial network of COVID-19 diffusion in Chinese Mainland by using multifractal scaling and measures. The spread of COVID-19 is a process, not a flash in the pan event, nor an unchangeable event. On the other hand, fractal is spatial order emerging from self-organized evolution (Hao, 2004). The fractal pattern is gradually developed from non-fractal structure (Benguigui *et al*., 2000). Therefore, it is necessary to generate the multifractal spectrums of COVID-19 spread on different dates. The COVID-19 dataset by prefecture-level cities and date is obtained from the real-time authorized reports announced by National Health Commission of the People's Republic of China and the provincial health commissions. We collect the cumulative number of confirmed cases in Chinese mainland from Jan 11 to Feb 29, 2020, which was during the early pandemic period of COVID-19 in China. There are a total of 50 consecutive days of spatial datasets, of which 42 days of data can be used for spatial analysis based on multifractal scaling. The COVID-19 datasets are matched with the latest administrative units of prefecture-level cities in ArcGIS 10.2. Thus, the spatial analysis database of COVID-19 is established for multifractal modeling.

The extraction and conversion of basic data of COVID-19 for fractal dimension calculation can be done using box-counting method, and the multifractal parameters of the spatial pattern can be calculated by least squares regression based on $\mu$-weight method. First of all, define a study area. That is, determine a study region for fractal measurement. This study area is applicable to all dates during our study period. We make a box nearly covering the whole range of Chinese mainland as our study area (Figure 2). This is the largest box, representing the first level box. The area of this box represents the measure area of multifractal network. Second, generate point set of data. Transform each prefecture-level administrative unit into a point according to its geographic center. Third, spatial disaggregation. Divide the box into four equal secondary boxes, and then divide each second level box into four third level boxes. In this way, recursively divide each box into four smaller boxes. The number of boxes increases exponentially and quickly reaches the limit of recursive subdivision of space. This method is termed functional box-counting (Chen and Wang, 2013; Lovejoy *et al*., 1987). Fourth, obtain datasets for estimating fractal dimension. Count the number of points in each box at each level, and produce the cumulative number of confirmed cases based on hierarchy of boxes. For each box at given level, we have a probability value of confirmed cases of COVID-19. Fifth, estimate the multifractal parameters. If the relationship between Renyi entropy of confirmed cases and the corresponding linear size of boxes satisfies logarithmic function, or if the relationship between generalized correlation function and the linear size of boxes follows scaling law, we can use the OLS-based regression method to calculate multifractal parameters by reconstructing probability measure (Chen and Wang, 2013).

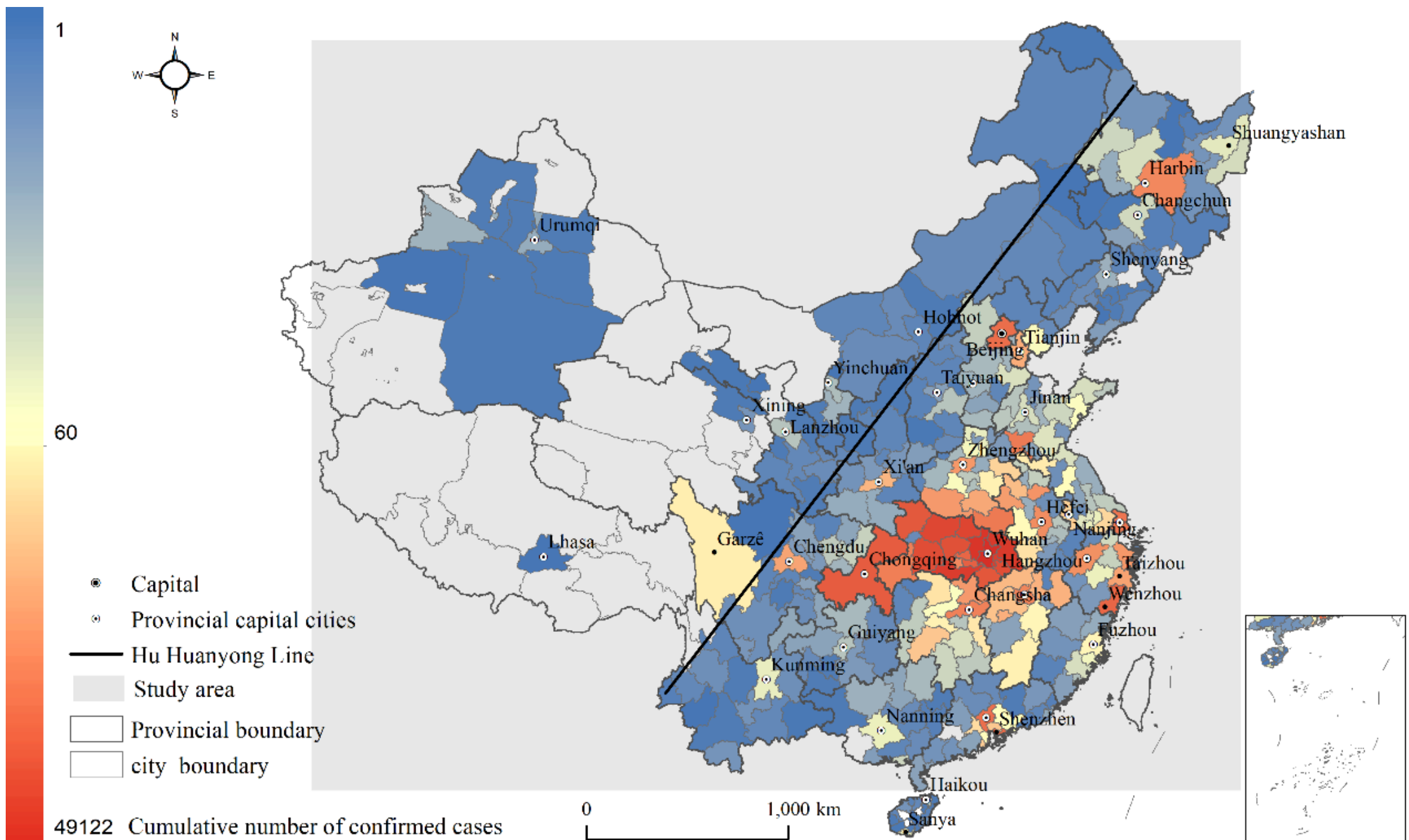

**Figure 2. Spatial distribution of the cumulative confirmed cases of COVID-19 in Chinese mainland** (up to Feb 29, 2020)

There are two ways to estimate multifractal parameters of complex spatial networks. One is estimate the global parameters first, and then use the Legendre transformation to work out the local parameters, and the other is to first calculate local parameters, and then use Legendre transformation to figure out global parameters. The first way: from global parameters to local parameter. The generalized correlation dimension can be directly computed by the linear regression based on equation (3). The independent variable is the logarithm of the linear size of the box, $\ln(\varepsilon)$, and the dependent variable is Renyi entropy, $I_q(\varepsilon)=(\ln\sum P_i(\varepsilon)^q)/(1-q)$. Specially, as shown above, for $q$=1, Renyi entropy should be replaced by Shannon entropy $I_1(\varepsilon)=-\sum P_i(\varepsilon)\ln P_i(\varepsilon)$. The regression coefficient gives the value of $D_q$, and the mass exponent $\tau(q)$ can be calculated with equation (4). Discretize the differential equation (7) yields a difference expression as follows

$$\alpha(q) \approx \frac{\Delta\tau(q)}{\Delta q}, \tag{9}$$

which can be utilized to estimate the singularity exponent $\alpha(q)$. Then the local fractal dimension $f(\alpha)$ can be computed with equation (8).

The second way: from local parameters to global parameters. This way relies heavily on what is called $\mu$-weight method. The $\mu$-weight method is actually a rescaling probability method. Reconstructing probability measure by normalizing the $q$th moment of probability $P_i(\varepsilon)$ as follows (Chhabra and Jensen, 1989)

$$\mu_i(\varepsilon) = P_i(\varepsilon)^q / \sum_{i=1}^{N(r)} P_i(\varepsilon)^q, \tag{10}$$

we have (Chhabra *et al*., 1989)

$$\alpha(q) = \lim_{\varepsilon \to 0} \frac{1}{\log \varepsilon} \sum_{i=1}^{N(\varepsilon)} \mu_i(\varepsilon) \log P_i(\varepsilon), \tag{11}$$

$$f(\alpha) = \lim_{\varepsilon \to 0} \frac{1}{\log \varepsilon} \sum_{i=1}^{N(\varepsilon)} \mu_i(\varepsilon) \log \mu_i(\varepsilon). \tag{12}$$

Then taking $\ln(\varepsilon)$ as an independent variable, and $\sum\mu_i(\varepsilon)\ln P_i(\varepsilon)$ or $\sum\mu_i(\varepsilon)\ln\mu_i(\varepsilon)$ as a dependent variable, we can estimate the values of $\alpha(q)$ and $f(\alpha)$ by OLS-based linear regression analysis (Chen, 2014a; Chen and Wang, 2013). Finally, the values of $D(q)$ and $\tau(q)$ can be converted by equations (4) and (8). Experience suggests that the second way is more convenient (Chen, 2014a). In this study, both ways were used, but the second way was the main one.

## 3. Results

### 3.1 Multifractal characteristics of the spatial pattern of COVID-19

The fractal development state of a complex system can be intuitively reflected by using double logarithmic plots, i.e., log-log plots. The log-log plots show the fractal features of the spatial pattern of COVID-19 diffusion (Figure 3), and its multifractal characteristics are revealed by the global and local fractal dimension spectrums (Figure 4). In fact, the spatial pattern of COVID-19 displays fractal property, but its fractal structure is not well-developed. This can treated as a type of weak multifractal phenomenon (Tarquis *et al*., 2017). The reason may be due to the government's lockdown measures. After all, fractal is an emergence pattern evolving from self-organized process (Hao, 2004). As seen in Table 1, the model has significant double logarithmic linear relation, indicating there is an obvious fractal nature of the spatial pattern of COVID-19 in the Chinese mainland.

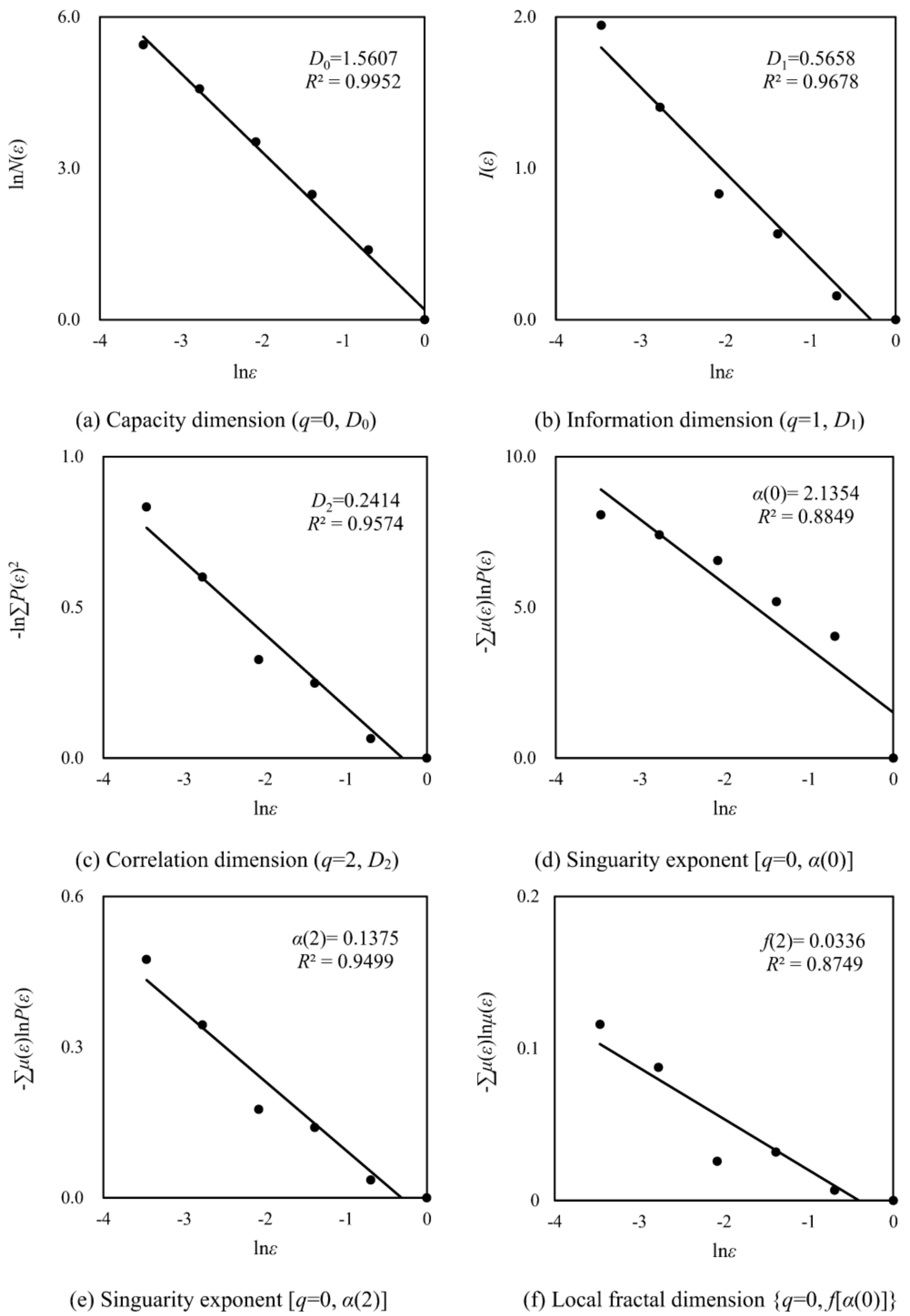

(a) Capacity dimension ($q$=0, $D_0$)

(b) Information dimension ($q$=1, $D_1$)

(c) Correlation dimension ($q$=2, $D_2$)

(d) Singuarity exponent [$q$=0, $\alpha(0)$]

(e) Singuarity exponent [$q$=0, $\alpha(2)$]

(f) Local fractal dimension {$q$=0, $f[\alpha(0)]$}

**Figure 3. The log-log plots of estimating multifractal parameters of the spatial pattern of COVID-19 on Feb 29, 2020 (examples)**

**Table 1 The capacity dimension, information dimension, and the correlation dimension of the spatial diffusion pattern of COVID-19 in Chinese mainland for every five days**

| Date | Capacity dimension $D_0$ | $R^2$ | Information dimension $D_1$ | $R^2$ | Correlation dimension $D_2$ | $R^2$ |
|---|---|---|---|---|---|---|
| **Jan 20** | 0.5369*** (0.0861) | *0.9068* | 0.1033** (0.0246) | *0.8151* | 0.0423** (0.0132) | *0.7211* |
| **Jan 25** | 1.4399*** (0.0656) | *0.9918* | 0.9960*** (0.0460) | *0.9915* | 0.5729*** (0.0288) | *0.9900* |
| **Jan 30** | 1.5321*** (0.0595) | *0.9940* | 0.9964*** (0.0649) | *0.9833* | 0.6052*** (0.0502) | *0.9732* |
| **Feb 5** | 1.5589*** (0.0551) | *0.9950* | 0.8561*** (0.0682) | *0.9753* | 0.4671*** (0.0486) | *0.9584* |
| **Feb 10** | 1.5598*** (0.0547) | *0.9951* | 0.7792*** (0.0654) | *0.9726* | 0.397*** (0.0420) | *0.9572* |
| **Feb 15** | 1.5607*** (0.0544) | *0.9952* | 0.6112*** (0.0556) | *0.9679* | 0.2714*** (0.0291) | *0.9560* |
| **Feb 20** | 1.5607*** (0.0544) | *0.9952* | 0.5852*** (0.0532) | *0.9680* | 0.2538*** (0.0269) | *0.9571* |
| **Feb 25** | 1.5607*** (0.0544) | *0.9952* | 0.5747*** (0.0526) | *0.9676* | 0.2472*** (0.0262) | *0.9569* |
| **Feb 29** | 1.5607*** (0.0544) | *0.9952* | 0.5658*** (0.0516) | *0.9678* | 0.2414*** (0.0255) | *0.9574* |

**Note:** The number of data points in each plot is 6, so the degree of freedom is 4. The parameter standard errors are quoted in parenthesis. *** significant at 1%; ** significant at 5%. Based on the significance level at 5%, fractal dimension plus or minus twice the standard error yields the margin of error, i.e., the lower and upper limits of fractal dimension.

According to the calculation results, the spatial pattern of COVID-19 diffusion in the Chinese mainland bears multifractal scaling properties. First of all, examine the basic parameters. The fractal parameters $D_0>D_1>D_2$ significantly. However, the log-log plots show that the fit of model to data is not so well when $q\neq0$, implying a poorly developed multifractal structure. Then, investigate the multifractal spectrums. The global and local multifractal parameters are calculated to obtain the corresponding generalized correlation dimension spectrum $D(q)$, and the local singularity spectrum $f(\alpha)$ (Figure 4). The $D(q)$ spectra appear as inverse S-shaped curves, rather than horizontal lines; and the $f(\alpha)$ spectrums are winding unimodal curves, instead of a single point. Both of them display typical signs of multifractal features. All in all, these calculations lend further support to the judgment that the spatial patterns of COVID-19 diffusion bear multifractal characteristics.

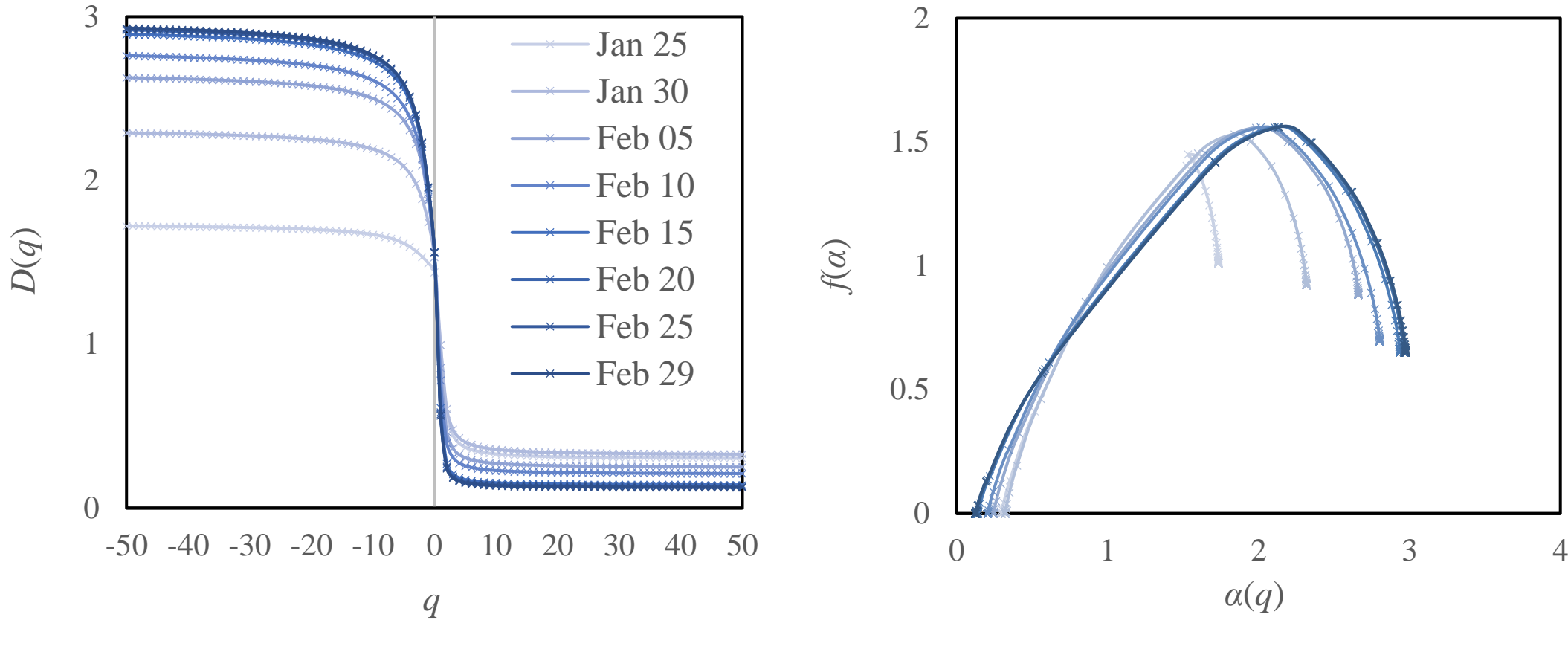

(a) The generalized correlation spectrums, $D(q)$ (b) The local singularity spectrums, $f(\alpha)$

**Figure 4. The multifractal spectrums of the spatial diffusion pattern of COVID-19 in Chinese mainland. Note:** Spatial information that cannot be seen on a map can be seen on multifractal spectral curves. The global spectral curves rises on the left and falls on the right over time. This means that the containment management in the diffusion centers is effective, but COVID-19 is still spreading on a small scale in marginal areas. The peak value of the local spectral curves slightly increases and the opening span slightly expands over time. This means that the total number of confirmed cases of COVID-19 is rising slightly, while the regional difference is increasing.

## 3.2 Multifractal spectrum analysis

The most intuitive reflection of geographic spatial diffusion is through maps. Using maps we can describe the laws of spatial diffusion and also reveal the features of spatial evolution (Morrill *et al*., 1988). By comparing different distribution maps of confirmed cases of COVID-19 in Chinese Mainland on different dates, we can see the characteristics of process and patterns of COVID-19's spread (Figure 5). However, the deep structure and dynamics of spatial diffusion of COVID-19 cannot be reflected or revealed through maps. Multifractals is a multi-scaling self-similar hierarchy with cascade structure. With the help of multifractal scaling, we can ignore specific geographical locations, project spatial information of density distribution into a hierarchical system of probability measurements, and then convert probability distribution into two sets of spectral curves. As mentioned above, one is the generalized correlation dimension spectrum, i.e., $D(q)$-$q$ spectrum or $D(q)$ curve, and the other is singularity spectrum, i.e., $f(\alpha)$-$\alpha(q)$ spectrum or $f(\alpha)$ curves. The former reflects global multifractal feature, and the latter mirrors the local multifractal structure. The combination of global and local multifractal spectrums can disclose the spatial diffusion characteristics of COVID-19 in the Chinese mainland.

The global dimension spectrum can be compared to a telescope. It reflects the overall characteristics of multifractal patterns. Let's firstly examine the global multifractal spectrums, namely $D(q)$ curves (Figure 4(a)). When $q \to +\infty$, the central regions with more infections would be brought into focus, especially in Wuhan and its surrounding areas, as well as some sub-central areas. Conversely, when $q \to -\infty$, the sparse regions with few infections will be highlighted, mainly in the edge areas or remote areas affected by COVID-19. In the $D(q)$ spectrum, the convergence value of $D(q)$ decreased slowly when $q>0$. This suggests that the spatial diffusion of COVID-19 in central regions was strictly confined by the lockdown measures. However, when $q \to -\infty$, the convergence values of $D(q)$ increased significantly over time, even exceeds the Euclidean dimension of embedding space ($d_E=2$). This implies that the spatial diffusion in edge areas was becoming so random and unpredictable that it was hard to control the epidemic in a short term. In short, the global multifractal spectrum curve rises on the left ($q<0$) and decreases on the right ($q>0$). The rising on the left means random diffusion in the edge area (low-density area with lower growth probability); while the decreasing on the right implies the effect of the strong lockdown in the central area (high-density area with higher growth probability).

The local multifractal spectrums reflect the micro mechanism of the spatial epidemic diffusion. In terms of function, the local dimension spectrum can be compared to a microscope. The $f(\alpha)$ curves display a strongly marked right-leaning unimodal curve, indicating the spatial pattern of COVID-19 takes on overall agglomeration characteristics (Figure 4(b)). There are two basic growth models of a multifractal system: if the $f(\alpha)$ spectrum is a right-leaning unimodal curve, it represents the growth pattern of spatial concentration (agglomeration); while if the $f(\alpha)$ spectrum takes on a left-leaning unimodal curve, it represents the growth pattern of spatial deconcentration (diffusion) (Chen, 2014a). It can be seen that the peak value the $f(\alpha)$ curve slightly rises and the curve opening span becomes larger over time. The increase of peak value means that the confirmed cases of COVID-19 gradually increase during the study period, while the larger opening means that the spatial diffusion leads to greater areal differences of COVID-19 distribution between high-density areas and low-density areas. In short, the $f(\alpha)$ spectrums reveal the spatial concentration pattern of COVID-19, indicating most of the epidemic infections are confined in a few regions, especially around Wuhan as well as cities closely connected with Wuhan. Thus, the timely control measures taken by Chinese government have made positive effects. Otherwise, if the $f(\alpha)$ spectrum is shown as a left-leaning unimodal curve, it implies that the spatial diffusion of COVID-19 would be too rapid to be effectively controlled. Fortunately, that is not the case. From the details of the local dimension spectrum, there is no significant variation, which also means that the spatial diffusion process of COVID-19 has been controlled to a certain extent.

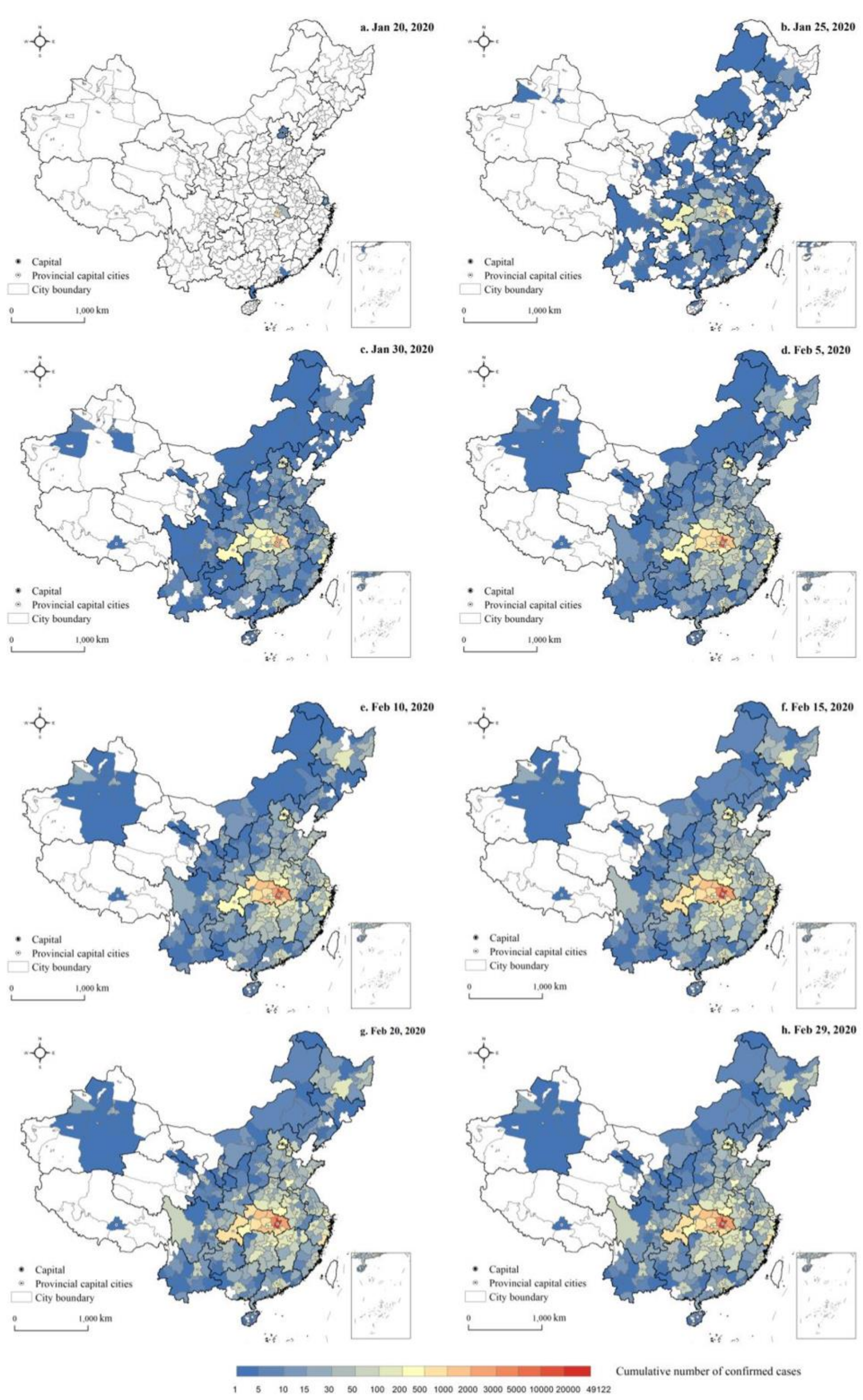

**Figure 5. The spatial distribution patterns of COVID-19 infections in Chinese mainland on different dates**

### 3.3 Stage characteristics of the spatial diffusion of COVID-19

The basis of this study is spatial data rather than time series. But the sample path of fractal parameters can be used for development stage analysis. A sample path is a subset taken from a time series (Diebold, 2007). The growth process and diffusion pattern represent two different sides of the same coin in geographical analysis. The growth in spatial diffusion over time usually presents as sigmoid curves (Banks, 1988; Morrill *et al*., 1988). The growth rate and acceleration behind the S-shaped curves can be used for stage division (Chen, 2014b). Let's examine three representative fractal dimensions: $D_0$, $D_1$ and $D_2$ (Figure 6(a) and Table 2). First, the capacity dimension $D_0$ roughly reflects the spatial diffusion pattern of COVID-19 from the perspective of space filling. The space-filling degree of infected cities in Chinese mainland increased quickly, and then the diffusion rate slowed down. Finally, it formed a constant spatial distribution, no longer spreading to new cities. Second, the information dimension $D_1$ reveals the spatial difference pattern. The spatial uniformity of COVID-19 infections firstly increased, and then the spatial difference increased gradually after Jan 26. Finally, the spatial distribution of epidemic infections remained stable. The significant decrease on Feb 12 is due to the statistical caliber adjustment of confirmed cases in Wuhan, but it has little effect on the overall trend. Third, the correlation dimension $D_2$ reveals the spatial dependence of epidemic infections among regions. The spatial correlation degree of the epidemic diffusion reached the strongest very quickly, then became weaker, and finally it was blocked effectively. Thus, the fractal dimension growth of spatial epidemic diffusion is different from that of cities. Urban growth is a natural process of self-organized evolution. So the fractal dimension curves of urban growth are continuous, which can be described by logistic function or quadratic logistic function (Chen, 2018).

**Table 2 The changing trend of three representative generalized correlation dimensions and their reflected geographic information**

| Parameter | Meaning | Trend | Inference |
|---|---|---|---|
| **Capacity** | Space-filling | Quasi S-shaped curve. | The spatial distribution range of |

| **dimension $D_0$** | degree | First rise, then tend to be stable | COVID-19 pandemic was expanded first and then converged |
|---|---|---|---|
| **Information dimension $D_1$** | Spatial uniformity | Firstly, it rose rapidly and then decreased gradually | The spatial differences of COVID-19 pandemic were first reduced and then enlarged |
| **Correlation dimension $D_2$** | Spatial dependence | Firstly, it rose rapidly and then decreased gradually | The spatial correlation of COVID-19 pandemic were first reduced and then enlarged |

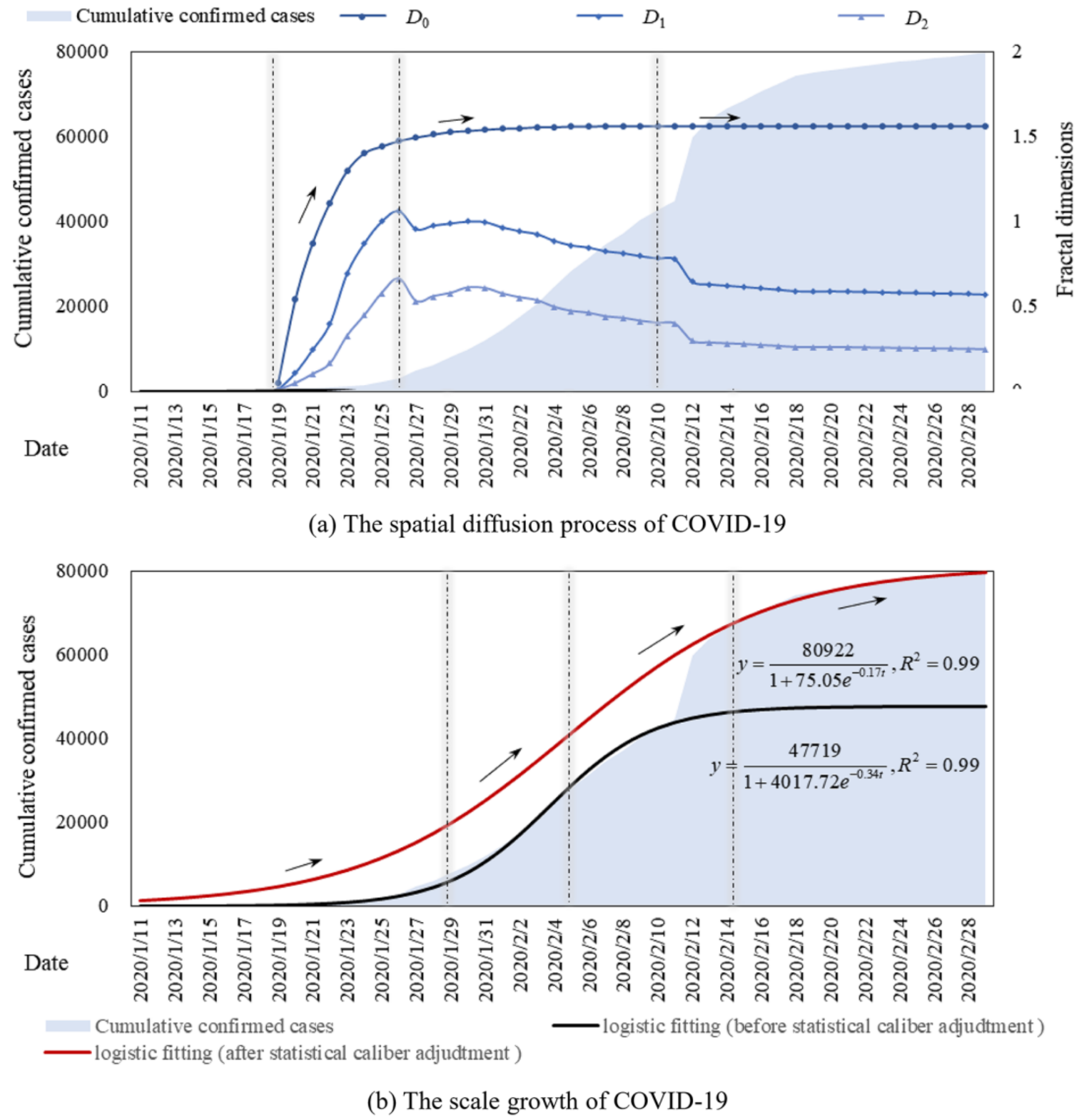

**Figure 6. The stages of the spatial diffusion and the scale growth of COVID-19. Note**: This is a study on spatial multifractal patterns. Due to the analyses are made on the multifractal patterns of 42 consecutive days, the fractal parameters forms a set of sample paths. A sample path is a subset of time series, representing a section taken from a time series. Time series is a theoretical concept, and empirical

analysis focuses on sample paths.

The process of spatial diffusion is not a uniform rate of change process. The changing curves of fractal dimensions reflect the stage characteristics of spatial diffusion of COVID-19 (Figure 6). The stage characteristics reveal the practical effects of prevention and control measures taken by Chinese government. The spatial distribution maps of COVID-19 infections only roughly display the spatial diffusion pattern, which is inconspicuous. In this regard, the changes of fractal dimensions play an irreplaceable role in the spatial analysis of epidemic diffusion. According to the trend of parameters $D_0$, $D_1$, $D_2$ (Table 2) and their corresponding spatial meanings, the spatial diffusion pattern of COVID-19 can be divided into 4 stages (Table 3). The four stages bear analogy with the urbanization curve which can be modeled by logistic function (Chen, 2014b).

①**The initial stage.** Before Jan 19, the epidemic spread inside Wuhan, showing a single point of burst. So the fractal dimensions are equal to 0.

②**The rapid diffusion stage.** From Jan 19 to 26, the space-filling degree ($D_0$), spatial uniformity ($D_1$) of the epidemic increased rapidly, and its spatial correlation degree ($D_2$) reached the highest. These suggest that the infected cities occupied Chinese mainland quickly in large-scale space. In this stage, the long-range diffusion of COVID-19 was the strongest. According to the first law of geography, geographical proximity affects the spatial correlation of regions. Spatial proximity may be the basis of the spatial diffusion of epidemic (Wang *et al*., 2020). However, action at a distance also play an important role in spatial diffusion (Chen *et al*, 2021). In addition to surrounding cities, the epidemic tended to spread from Wuhan to remote cities, including megacities (e.g., Shenzhen, Beijing, Shanghai), provincial capital cities, and some southeastern coastal cities (e.g., Wenzhou, Taizhou). These cities boast a developed economy and are the main nodes of urban economic connections. The remote spatial diffusion of COVID-19 may be dominated by economic connections with high liquidity such as tourism and business travel (Shi and Liu, 2020).

③**The deceleration diffusion stage.** From Jan 27 to Feb 9, the diffusion rate of infections decreased gradually, and the spatial difference increased. This implies that though more cities were infected, their infections grew slower. While in central regions, their infections grew faster significantly. Besides, the spatial correlation became weaker, implying the epidemic mainly spread to contiguous cities from regional central cities. So it can also be described as the hierarchical diffusion stage, dominated by contagious diffusion. Hierarchical diffusion is a common pattern in the process of geographic spatial diffusion (Morrill *et al*., 1988). As a whole, the spatial diffusion of COVID-19 infections is seemly bounded by China's population distribution boundary, i.e., Hu Huanyong Line (Figure 2). This line delineates the striking difference in the distribution of China's population (Hu, 1935). As seen in Figure 2 and Figure 5, the epidemic infections in Chinese mainland are mainly located in the east of 'Hu Huanyong Line', which shares high similarity with the spatial pattern of population distribution in China.

④**The contraction stage.** After Feb 10, the spatial diffusion pattern of COVID-19 reached a relatively stable state. The value of $D_0$, $D_1$ and $D_2$ gradually converged, reflecting the spatial contraction of epidemic diffusion. About two weeks after the Wuhan lockdown (Jan 23), the spatial diffusion trend of COVID-19 has been effectively controlled in Chinese mainland. Besides, the cooperation mechanism on joint prevention and control among regions further blocked the contagious diffusion. This suggests the isolation over regions and reducing inter-regional movement had basically curbed the spread of COVID-19 in Chinese Mainland.

Corresponding to the spatial diffusion process of COVID-19, the scale growth of the cumulative number of confirmed cases also presents stage characteristics. The sample paths of cumulative confirmed cases in Chinese Mainland and large cities take on sigmoid curves (Figure 6(b)). Studies have found that the growth of cumulative confirmed cases can be modeled by logistic function (Chen *et al*, 2021; Consolini and Materassi, 2020; Martelloni and Martelloni, 2020; Pelinovsky *et al.*, 2020; Wang *et al.*, 2020). In this study, two logistic functions are employed to model the scale growth of COVID-19 infections from Jan 11 to Feb 29, due to the statistical caliber adjustment of diagnostic criteria. According to their growth rate and acceleration, the scale growth of COVID-19 can also be divided into 4 stages: the initial stage (before Jan 19), the acceleration stage (Jan 19-Feb 5), the deceleration stage (Feb 6-14), and the late stage (after Feb 14). It can be found that the increase of new confirmed cases reached a peak point on Feb 5. After that, the growth rate slowed down and remained stable in late February. While the spatial diffusion of COVID-19 has reached its peak stage and stable stage earlier. Thus it can be seen that the development of the spatial diffusion process is faster than the scale growth process of COVID-19. On the whole, the stage of spatial diffusion of COVID-19 reflected by fractal parameter series is slightly different from that reflected by time series of cumulative number of confirmed cases (Figure 6, Table 3). The stage changes reflected by the fractal dimension sequence are more profound, detailed, and specific. The two types of time series, namely cumulative confirmed cases and fractal parameters, can reflect the spatial filling in the diffusion process of COVID-19, but the time series of cumulative confirmed cases cannot reveal the spatial differences and correlations in this process.

**Table 3 Different stages of spatial diffusion process and scale growth process of COVID-19 in Chinese mainland**

| Stages | Spatial diffusion | Periods | Scale growth | Periods |
|---|---|---|---|---|
| **Stage 1** | Initial stage | Before Jan 19 | Initial stage | Before Jan 19 |
| **Stage 2** | Rapid diffusion stage | Jan 19 - 26 | Acceleration stage | Jan 19 - Feb 5 |
| **Stage 3** | Hierarchical diffusion stage | Jan 27 - Feb 9 | Deceleration stage | Feb 6 - 14 |
| **Stage 4** | Contraction stage | After Feb 10 | Late stage | After Feb 14 |

## 4. Discussion

The above calculations show that the spatial network of COVID-19 infections in Chinese mainland takes on multifractal characteristics. The distribution pattern and diffusion process of COVID-19 can be characterized by multifractal spectrums. The key points of the analyzed results are as follows (Table 4).Firstly, the spatial pattern of COVID-19 in Chinese mainland is mainly characterized by global aggregation and local diffusion. Though the epidemic infections spread nationwide, few cities were hit severely. In the global multifractal dimension spectrum, the convergence value of $D(q)$ increased significantly over time when $q<0$, suggesting the local diffusion of COVID-19 infections (Figure 4(a)). However, the local singularity spectrum $f(\alpha)$ always takes on a right-leaning unimodal curve, implying a global aggregation pattern (Figure 4(b)). The epidemic infections in central regions were severe significantly than in marginal regions. Secondly, the spatial diffusion pattern of COVID-19 in Chinese mainland show stage characteristics, suggesting the positive effects of prevention and control measures taken by the government. According to the fractal dimension growth curves, it can be divided into 4 stages: the initial stage, the rapid diffusion stage, the hierarchical diffusion stage and the contraction stage, from the single point of burst, multi-points spread to the overall outbreak, and finally full containment. And the spatial correlation degree decreased with time. In general, the spatial diffusion of COVID-19 has been contained in a short term. The rapid diffusion stage is the transition period of epidemic diffusion. A series of spatial isolation measures implemented in China has played an important role in the control of COVID-19.

**Table 4 Main points of fractal-based spatial analysis for COVID-19 pandemic in Chinese mainland**

| Item | Feature | Inference |
|---|---|---|
| **Log-log plot** | Straight-line trend | Fractal pattern |
| **Fractal parameters** | $D_0>D_1>D_2$ | Multifractal scaling |
| **Global multifractal spectrum, $D(q)$ curve** | Inverse S-shaped curve | Multifractal diffusion |
| **Local multifractal spectrum, $f(\alpha)$ curve** | Right-leaning unimodal curve | Global aggregation, and local diffusion |
| **Fractal dimension growth curves** | Quasi S-shaped curves | Prevention and control measures are effective |

Multifractal scaling has two major application directions in complexity exploration. One is to study nonlinear dynamical processes, and the other is to study spatial patterns of scale-free heterogeneous distributions. The former is based on time series (e.g., Ihlen, 2013; Jiang *et al*., 2019; Kantelhardt *et al*., 2002; Tarquis *et al*., 2017; Tessier *et al*., 1996), while the latter relies heavily on cross-sectional datasets (e.g., Appleby, 1996; Chen, 2014a; Chen and Wang, 2013; Huang and Chen, 2018; Long and Chen, 2017; Murcio *et al*., 2015; Song and Yu, 2019). Therefore, there are two categories of multifractals studies in literature: one is multifractal time series analysis, and the other is multifractal spatial analysis (Table 5). This paper belongs to the second category, aiming at exploring the spatial complexity of the spread of COVID-19 by using multifractals. Based on the spatial analysis results of consecutive days, the evolution characteristics of COVID-19 diffusion are also revealed by fractal dimension time series. Multifractal spatiotemporal analysis is one of the features of this work. However, the sample paths of confirmed cases of COVID-19 in Chinese mainland do not support multifractal time series analysis due to the following reasons. First, the sample paths we obtained are too short; second, data caliber has been adjusted midway; third, the sequences have definite S-shaped growth curve trend, and this trend suggests no multifractal structure in the time series (Figure 6(b)).

**Table 5 A comparison between two types of multifractal studies: multifractal time series analysis and multifractal spatial analysis**

| **Type** | **Temporal multifractal analysis** | **Spatial multifractal analysis** |
|---|---|---|
| **Research object** | Temporal process | Spatial distribution |
| **Data** | Time series data | Spatial datasets |
| **Method** | Wavelet Transform (WT), De-trended Fluctuation Analysis (DFA); Phase Space Reconstruction (PSR), etc. | Box-counting method (BCM); Sandbox method (SBM); Fixed radius method (FRM); Fixed mass method (FMM), etc. |
| **Parameters** | Global and local parameters | Global and local parameters |
| **Scaling** | Temporal scaling | Spatial scaling |
| **Space** | Phase space | Real space |

Epidemic spatial diffusion and its dynamics analysis is an important subject of geographical spatial analysis, and has been studied for many years. Most of these studies have described the spatio-temporal pattern of the epidemics by spatial autocorrelation analysis and kernel density estimating (Meng *et al*., 2005; Wang *et al*., 2008; Fang *et al*., 2009; Kang *et al*., 2020). Nevertheless, a critical problem with these methods is their sensitivity to the spatial scale (Negreiros *et al*., 2010; Zhang *et al*., 2019). The novelty of this work lies in that it reveals the spatio-temporal characteristics of COVID-19 plague transmission in Chinese Mainland by means of multifractal scaling analysis. The spatial diffusion of COVID-19 is a complex process, showing irregular and scale-free features. Given this, fractal analysis is a powerful tool for spatio-temporal modeling of spatial epidemic diffusion from the perspective of scale-free analysis. Multifractal scaling analyses provide two sets of parameters and spectrums, forming a complete understanding of the spatial diffusion pattern of COVID-19 from general to detail. Our research is not yet complete and there is great room for improvement. The shortcomings of this study are as below: first, our analysis is based on the COVID-19 dataset from real-time authorized reports, so that the incubation period is not considered.

But it has little impact on the overall trend of epidemic evolution. Second, the log-log plots display a poor-developed fractal structure of COVID-19, which may be caused by the data quality. This study takes the prefecture-level cities as basic spatial units, so that the scaling range for fractal dimension estimation is short. On the other hand, the spatial epidemic diffusion is not bounded by the administrative boundary. If there is more accurate location datasets of the confirmed cases, including county-level city or even towns, further detailed spatial analysis can be performed. What is more, the lack of in-depth time series analysis is also a deficiency of this article.

## 5. Conclusion

Modeling the spatial diffusion process and pattern of COVID-19 in Chinese mainland can provide valuable experience for future theoretical and positive studies, and for further improvement of future public health system emergencies. In this work, multifractal scaling model is employed to characterize the spatial diffusion pattern of COVID-19 pandemic, based on the cumulative number of confirmed cases from the authority reports. The main conclusions can be drawn as follows. **First, the spatial pattern of COVID-19 in Chinese mainland bears multifractal characteristics.** This suggests that the spread of the COVID-19 epidemic depends on the urban system, which has been proved to have a multifractal structure. Moreover, the spatial distribution pattern of human population may be the geometric support of the multifractal distribution of COVID-19. Population takes on multifractal distribution in geographical space. Most of the population is located in the southeastern part of China. Accordingly, most of the COVID-19 pandemic are located in the southeast China. The spatial diffusion pattern of COVID-19 has well coincidence relations with the spatial pattern of population in China, both in the geographical distribution and structural hierarchy. **Second, the spatial pattern of COVID-19 in Chinese mainland is characterized by global clustering with local diffusion.** This suggests that China has effectively contained the COVID-19 pandemic in a short term and has significant performance in this epidemic prevention and control. In terms of multifractal spectrums of local parameters, large-scale infections were confined in central regions, i.e., Wuhan city and its surrounding region. The epidemic situation in other regions was in a controllable state. However, the multifractal spectrums of global parameters shows that there was small-scale disordered diffusion of COVID-19 pandemic in lots of local places during this study period. **Third, the spatial diffusion process of COVID-19 in Chinese mainland fell into 4 stages.** The fractal dimension growth processes take on sigmoid curves. The capacity dimension growth curve reflects the change of space filling process of COVID-19, the information dimension growth curve reflects the change of spatial difference of COVID-19, and the correlation dimension growth curve reflects the change of spatial dependence of COVID-19. The space filling curve increased first and became stable rapidly. The spatial difference and dependence curves increased first and then decreased and finally become stable. By the three curves, the spatial diffusion process of COVID-19 can be divided into four stages: initial stage, rapid diffusion stage, hierarchical diffusion stage, and the contraction stage.

## Acknowledgments

This research was sponsored by the National Natural Science Foundation of China (Grant No. 41671167). The support is gratefully acknowledged.